\documentclass[useAMS,usenatbib]{mn2e}
\usepackage{graphicx}
\usepackage{aas_macros}
\usepackage{color}

\title[Star Formation in the Smith Cloud]{A Search for Star Formation in the Smith Cloud}
\author[Stark et al.]{David V. Stark$^{1}$\thanks{dstark@email.unc.edu}, Ashley D. Baker$^{1}$, Sheila J. Kannappan$^{1}$\\
$^{1}$Physics and Astronomy Department, University of North Carolina, Chapel Hill, NC 27516}

\begin{document}


\date{Accepted 2014 October 16. Received 2014 October 16; in original form 2014 June 20}

\pagerange{\pageref{firstpage}--\pageref{lastpage}} \pubyear{2014}

\maketitle

\label{firstpage}

\begin{abstract}
Motivated by the idea that a subset of HVCs trace dark matter
substructure in the Local Group, we search for signs of star formation
in the Smith Cloud, a nearby $\sim 2 \times 10^6 \rm{M}_{\odot}$ HVC
currently falling into the Milky Way.  Using $GALEX$ NUV and
$WISE$/2MASS NIR photometry, we apply a series of color and apparent
magnitude cuts to isolate candidate O and B stars that are plausibly
associated with the Smith Cloud.  We find an excess of stars along the
line of sight to the cloud, but not at a statistically significant
level relative to a control region.  The number of stars found in
projection on the cloud after removing an estimate of the
contamination by the Milky Way implies an average star formation rate
surface density of $10^{-4.8\pm0.3}$~M$_{\odot}$~yr$^{-1}$~kpc$^{-2}$,
assuming the cloud has been forming stars at a constant rate since its
first passage through the Milky Way $\sim$70~Myr ago.  This value is
consistent with the star formation rate expected based on the average
gas density of the cloud.  We also discuss how the newly discovered
star forming galaxy Leo P has very similar properties to the Smith
Cloud, but its young stellar population would not have been detected
at a statistically significant level using our method.  Thus, we
cannot yet rule out the idea that the Smith Cloud is really a dwarf
galaxy.
\end{abstract}


\section{Introduction}
\label{sec:intro}

Since the discovery of high velocity clouds (HVCs), astronomers have
explored a variety of scenarios to explain their nature and origin
(see \citealt{Wakker97} and references therein).  As a possible
solution to the `missing satellites problem' (e.g.,
\citealt{Klypin99}), both \citet{Blitz99} and \citet{Braun99} proposed
that HVCs, or the subset of compact HVCs (CHVCs), trace low mass dark
matter halos in the Local Group.  \citet{Sternberg02} argue against
CHVCs being embedded in halos at typical distances of $\sim$1 Mpc
because their dark matter halos would have unrealistic densities,
although they find that CHVCs are consistent with being embedded in
dark matter minihalos that are circumgalactic with distances
of $\sim$150 kpc.  More recently, a new class of HVCs, called
ultra-compact HVCs (UCHVCs), was discovered in the ALFALFA survey and
have properties consistent with being dark matter halos spread
throughout the Local Group \citep{Giovanelli10,Adams13}.

If all or a subset of HVCs trace dark matter substructure around the
Milky Way, they might contain some (potentially very faint) stellar
population.  In contrast, in the absence of dark matter, such star
formation may never occur.  Based on calculations without dark matter,
\citet{Christodoulou97} argue that the dense cores in high velocity
clouds never reach high enough mass to collapse on their own, and the
cloud-cloud collision timescale for these cores within a single
complex is $>$~1~Gyr.  However, simulations of infalling HVCs without
dark matter imply they survive less than $\sim$100~Myr before they are
dispersed \citep{Heitsch09,Nichols09,Joung12}.


  Aside from considerations about dark matter content, star
  formation in HVCs faces additional challenges.  The average observed
  HI column density in HVCs is roughly $\sim$10$^{19}$ cm$^{-2}$
  \citep{Putman12}, whereas star formation (at least in galaxy disks)
  is typically associated with HI column densities above 10$^{21}$
  cm$^{-2}$ (e.g., \citealt{deBlok06}).  Molecular gas, which is
  considered direct fuel for star formation, has only been
  successfully detected in a handful of HVCs
  \citep{Richter01,Sembach01,Hernandez13}.  In addition, HVCs have
  very low metallicity (10-30\% solar) and dust content. No HVCs, with
  the exception of the Magellanic Stream, show depletion of refractory
  elements, and dust has only been tentatively detected from FIR
  emission in a handful of cases \citep{Putman12}.  Low dust content
  may be an important factor when considering star formation since
  H$_2$ formation is catalyzed by dust grains \citep{Gould63}.
  However, alternative viewpoints say that the formation of H$_2$ in
  itself is not the cause of star formation but simply a process that
  also occurs in dense gas \citep{MacLow12}, so the existence of dust
  in itself may not be crucial.  The recent discovery of star
  formation in low gas density, low metallicity XUV disks
  \citep{Thilker07} calls into question many presumptions about the
  conditions in which star formation can occur.

There have been multiple attempts to search for stellar populations in
high velocity clouds.  An extensive search of POSS imaging by
\citet{Simon02} found no associated stellar populations in a sample of
250 northern sky HVCs, suggesting that HVCs are not associated with
normal but faint dwarf galaxies like those already found in the Local
Group.  Several smaller studies have focused on identifying
overdensities of red giant or blue luminous stars at the positions of
high-latitude and often compact HVCs, and these studies have
likewise found no associated stellar populations
\citep{Willman02,Hopp03,Siegel05,Hopp07}. Based on their own search
for stellar populations associated with HVCs in combination with those of
other groups up to that point, \citet{Hopp07} argued that no more than
4 per cent of HVCs could harbor star formation.

On the other hand, some searches for stellar populations associated
with HVCs have proven fruitful.  A single YSO in close proximity to a
high-velocity HI emission peak was identified by \citet{Ivezic97}, who
argued that the probability of such an occurrence being observed due to
a random projection is extremely low.  More recently,
\citet{McQuinn13} and \citet{Rhode13} reported on a stellar population
associated with Leo P, an object originally classified as an ultra
compact HVC (UCHVC) after its discovery in the ALFALFA survey
\citep{Giovanelli13}.

In this study, we search for signs of star formation in Complex GCP,
commonly known as the Smith Cloud \citep{Smith63}, in an
  attempt to constrain the likelihood that it hosts a dark matter
  halo.  The Smith Cloud is a nearby low latitude HVC with total
  neutral and ionized hydrogen mass $>2\times 10^6$~M$_{\odot}$
  \citep{Lockman08,Hill09}.  There are several advantages to studying
  the Smith Cloud.  It has a peak column density of $\sim 5
  \times10^{20}$~cm$^{-2}$, putting it in a low gas density regime
  that still hosts star formation as seen by \citet{Bigiel08},
  although we note that the Smith Cloud is dynamically very different
  from the extended galactic HI disks in that work.  While the low
  latitude of the Smith Cloud has the practical disadvantage of higher
  foreground extinction, its ongoing interaction with the Milky Way
  may trigger star formation (as in, e.g.,
    \citealt{Casetti14}).

The fact that the Smith Cloud has survived as long as it has is
additional evidence that it is embedded within a dark matter halo.
Extrapolation of its trajectory implies it must have already passed
through the Milky Way disk $\sim$70~Myr ago and is on track to collide
with the disk again in another $\sim$30~Myr
\citep{Lockman08}. \citet{Nichols14} argue it could never have
survived its passage through the Galactic disk without dark matter,
although the recent detection of magnetic fields by \citet{Hill13}
provides an additional means to contain the cloud.  A final advantage
of studying the Smith Cloud is that it is one of the few HVCs with a
well determined distance based on multiple techniques
\citep{BlandHawthorn98,Putman03,Lockman08,Wakker08}, allowing an
additional means to isolate stars that could be associated with the
cloud.  Detection of a stellar population associated with the Smith
Cloud could suggest that some HVCs are indeed associated with dark
matter halos/subhalos, and perhaps only HVCs with such halos are
capable of star formation.

In this work, we use a combination of $GALEX$ NUV and 2MASS/$WISE$ NIR
photometry to search for evidence of recent star formation in the
Smith Cloud.  To accomplish this goal, we apply a series of color and
magnitude cuts to isolate young stars consistent with being associated
with the Smith Cloud.  While we find a slight overdensity of young
stars in projection in the region of the Smith Cloud, this overdensity
is not statistically significant compared to the density of a control
region.  However, our estimate of the star formation rate is
consistent with what would be expected based on the observed gas
density.  We cannot rule out the possibility that the Smith Cloud
hosts low level star formation, and our analysis suggests that the
recently discovered stellar population in Leo P, whose global
properties are similar to the Smith Cloud, would not have been
detected at a statistically significant level by our methods.

\section[]{Data and Methods}
\label{sec:data}

Below we describe our sources of data.  All reported magnitudes
are corrected for foreground extinction using the dust maps of
\citet{Schlegel98} and the extinction law of \citet{Odonnell94}.

\subsection{21cm Data}
\label{sec:21cmdata}

\begin{figure}
\includegraphics[width=84mm]{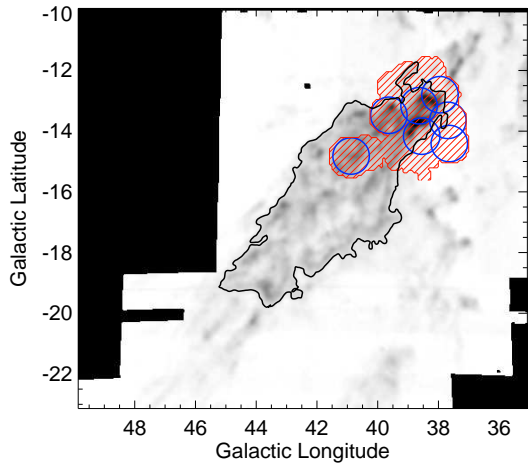}
\caption{HI column density map of the Smith Cloud, adapted from
  \citet{Lockman08}.  Square black regions represent masked
  pixels in the data cube.  The thick black line shows the outermost
  N$_{\rm HI}=5\times 10^{19}$~cm$^{-2}$ contour used to define the
  edge of the cloud.  The red filled region illustrates the extent of
  the $GALEX$ data used in this study, with the positions of deeper
  $GALEX$ fields overlaid in blue.}
\label{fig:galexfields}
\end{figure}

This study uses GBT 21cm observations of the Smith Cloud presented in
\citet{Lockman08}.  To separate Smith Cloud emission from foreground
Milky Way emission, a zero-moment map was created using all channels
corresponding to V$_{LSR}$=75--125~km~s$^{-1}$.  For our analysis, we
need to define what is considered to be `on' versus `off' the
cloud.  The outermost N$_{\rm HI}=5\times 10^{19}$~cm$^{-2}$ contour
traces the overall head-tail structure of the cloud (see
Fig.~\ref{fig:galexfields}), so we refer to all regions within it as
`on' the cloud, while regions beyond it are considered `off' the
cloud.  In \S~\ref{sec:likelihood}, we address how our results are
affected by this definition.

\subsection{Photometric Data}
\label{sec:photodata}

\subsubsection{$GALEX$}
\label{sec:galex}

Deep $GALEX$ NUV imaging was obtained for 7 fields in the vicinity of
the Smith Cloud (program GI6-041, PI Stark), with most fields covering
regions on or around the cometary head where the densest gas is
located.  Each field covers a circular diameter of 1.2$^{\circ}$ and
integration times vary from 1800--4697~seconds.  Due to the failure of
the FUV channel on $GALEX$ prior to our observing program, no FUV imaging is
available for these same regions.  Additional All-sky Imaging Survey
(AIS) data have been used to enlarge the NUV coverage beyond the
border of the cloud.  While the AIS images only have typical exposure
times of $\sim$100~seconds, they are still deep enough to detect young
OB stars at the distance of the Smith Cloud (see
\S~\ref{sec:magcuts}).  In total, the $GALEX$ coverage extends over
11.04~deg$^2$, with 6.08~deg$^2$ and 4.96~deg$^2$ lying `on' and
`off' the cloud, respectively, using the definitions from
\S~\ref{sec:21cmdata}.

In addition, we acquired archival data for three Medium Imaging Survey
(MIS) depth fields (totaling 3.39~deg$^2$) at comparable Galactic
latitude but opposite Galactic longitude to the Smith Cloud.  We use
these as control regions far from the Smith Cloud but roughly
equidistant from the Galactic bulge and mid-plane.  The central coordinates of
these three fields are given in Table~\ref{table:offfields}.

All data products used in this study result from the standard $GALEX$
imaging pipeline \citep{Morrissey07}.

\begin{table}
\caption{Control Region Fields}
\label{table:offfields}
\begin{tabular}{@{}ccc}
\hline
Field & $l^{\circ}$ & $b^{\circ}$ \\
\hline
1 & 317.49 & -12.22 \\
2 & 320.25 & -14.91 \\
3 & 326.01 & -14.25 \\
\hline
\end{tabular}
\end{table}


\subsubsection{WISE \& 2MASS}
\label{sec:nir}

We have taken $WISE$ 3.6$\mu$m and 4.5$\mu$m photometry\footnote{ We
  explored the use of $W3$ and $W4$ imaging to look for heated dust
  emission in the Smith Cloud that would be characteristic of young
  stars.  However, the IR emission from the foreground Milky Way
  dominates the field.  Disentangling the two emission sources is not
  trivial, so we did not pursue this further.} from the $WISE$ All-Sky
Source Catalog \citep{Wright10}.  After rejecting extended/blended
objects and sources contaminated by artifacts, the $GALEX$ and $WISE$
data are cross-matched using a 3\arcsec match radius.
  Approximately 45 per cent of the $GALEX$ sources lack detected
  $WISE$ counterparts; the majority of these sources have extremely
  low signal-to-noise ($<$3) and lie near the sensitivity limit of the
  $GALEX$ data.  We ignore these in our subsequent analysis since they
  are inconsistent with being OB stars at the distance of the Smith
  Cloud (see \S~\ref{sec:magcuts}).

We also use 2MASS $JHK$ photometry, which has already been
crossmatched with $WISE$ sources and included in the $WISE$ All-Sky
Source Catalog. Because 2MASS data are shallower than $WISE$ data, a
small fraction of $GALEX$ objects with corresponding $WISE$ data lack
2MASS counterparts.  Therefore, $GALEX$ and $WISE$ photometry are used
in our primary analysis, with 2MASS being incorporated when available.

In the Smith Cloud fields (including regions both on and
off the cloud), there are a total of 41,783 objects with measured
$GALEX$ NUV and $WISE$ $W1$ and $W2$ magnitudes.  Approximately 90 per cent of
these have measured 2MASS $JHK$ magnitudes.  In the control field,
there are a total of 18,070  sources with NUV, $W1$, and $W2$ magnitudes, and
again $\sim$90 per cent also have measured 2MASS magnitudes.

\subsection{Identification of Candidate Stars}
\label{sec:obid}

\subsubsection{Synthetic Stellar Spectra}
\label{sec:synthetic}

We use synthetic stellar spectra to determine colors and apparent
magnitudes consistent with OB stars at the distance of the Smith
Cloud.  The BaSeL 2.2 library \citep{Lejeune98} provides synthetic
spectra that cover the necessary wavelength range to include all
$GALEX$, $WISE$, and 2MASS passbands, and the models range over
stellar surface temperatures from 2000 K to 50000 K.  We choose
stellar models with metallicities of 0.05Z$_{\odot}$ and
0.5Z$_{\odot}$, which bracket the range of possible metallicities of
the Smith Cloud (0.05--0.4Z$_{\odot}$) constrained by \citet{Hill09}.
Model magnitudes are obtained by convolving the synthetic spectra with
the filter profiles of all passbands.

\subsubsection{Color Cuts}
\label{sec:colorcuts}

\begin{figure*}
\includegraphics[width=170mm]{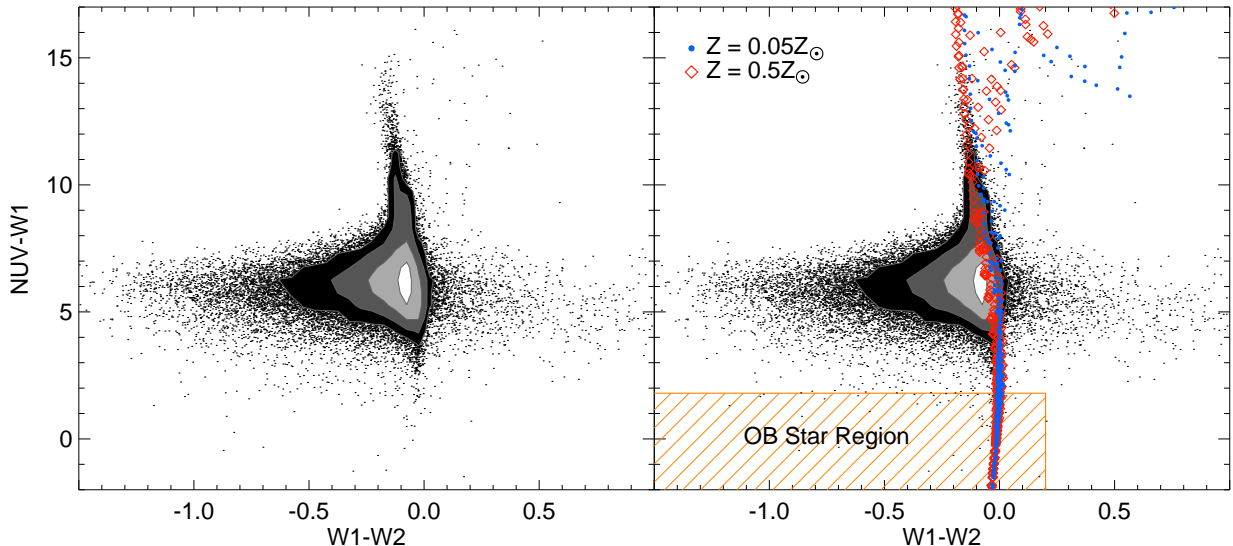}
\caption{{\it(left)} Our $GALEX$/$WISE$ crossmatched sources, both on
  and off the cloud.  Filled contours represent regions where the
  number of stars in a 0.5$\times$0.05 box (NUV-$W1 \times W1-W2$)
  falls above 100, 250, 500, and 1000. {\it (right)} Model spectral
  colors (red and blue points) from the BaSeL 2.2 library overlaid on
  top of the data.  The two BaSeL model sets shown, with metallicities
  of 0.05Z$_{\odot}$ and 0.5Z$_{\odot}$, approximate the minimum and
  maximum possible metallicities of the cloud (0.05Z$_{\odot}$ and
  0.4Z$_{\odot}$; \citealt{Hill09}).  The hashed region defined by
  NUV-$W1<1.8$ and $W1-W2<0.2$ shows the expected positions of OB
  stars free of contamination from QSOs and YSOs.  The synthetic
  colors derived from both models are almost identical in the region
  of parameter space occupied by OB stars.  The scatter in $W1-W2$ at
  higher NUV$-W1$ is real (\S~\ref{sec:synthetic}) and does not
  compromise our analysis.}
\label{fig:synthetic}
\end{figure*}

To isolate OB stars, we explore how the BaSeL spectral models are
distributed in an NUV$-W1$~vs.~$W1-W2$ diagram
(Fig.~\ref{fig:synthetic}). In the synthetic models, all OB stars
(defined to have surface temperatures $>10^4$~K) have NUV$-W1 < 1.8$,
which we use as our first cut.  We also find that the model stars have
$W1-W2\sim0$, which is consistent with prior studies by \citet{Wu12}
and \citet{Yan13}.  These same authors show that quasars
  become a significant number of point sources at $W1-W2 > 0.2$, while
  \citet{Mace13} find YSOs at $W1-W2 > 0.5$.  Thus, to avoid
  contamination from these point sources while keeping nearly all
  stars, we keep only objects with $W1-W2 < 0.2$.  We place no lower
  limit on the $W1-W2$ colors of stellar candidates since we find that
  stars tend to be biased towards bluer colors at dimmer magnitudes
  (see \S~\ref{sec:scob}).  Wherever 2MASS magnitudes are available,
we further isolate OB stars by rejecting all sources with $J-H > 0 $
and $H-K > 0.02$.  These cuts are taken from \citet{Straizys09}, who
measured these colors for stars of known spectral type.
  For all three of our color cuts, we allow a star to pass a color cut
  if its 1$\sigma$ uncertainty potentially places it in OB star
  parameter space, even if its measured value is outside.

Fig.~\ref{fig:synthetic} shows the model colors overlaid on our data
for the Smith Cloud $GALEX$ fields (both on and off the cloud).  The
model colors are consistent with the data, although there is
considerable scatter in $W1-W2$ at $2<$NUV$-W1<9$.  Such scatter has
been seen in other studies \citep{Wu12,Yan13} and does not affect our
analysis, which is restricted to lower values of NUV$-W1$.  The
crosshatched area denotes the region of parameter space occupied by OB
stars.

\subsubsection{Apparent Magnitude Cuts}
\label{sec:magcuts}

Since the Smith Cloud has a well constrained distance of 12.4$\pm$1.3
kpc \citep{Lockman08}, we have the advantage of being able to predict
the range of possible apparent magnitudes of its OB star population.
Expected NUV, $W1$, and $J$ band absolute magnitudes are estimated
using BaSeL spectral models to estimate surface flux while adopting
typical stellar radii as a function of spectral type
\citep{Habets81,Massey04,Repolust04,Massey05,Massey09}.  We estimate
that between surface temperatures of 1-5$\times10^4$~K, OB star
luminosities can vary by a factor of $\sim$2$\times 10^4$, $\sim$500,
and $\sim$700 in the NUV, $W1$, and $J$ bands respectively.  There are
additional variations in apparent OB star brightness caused by stars
occupying the full width of the cloud along the line of sight, which
we estimate to be approximately 325~pc (1.5~degrees), which was
measured in the region where we have imaging and assumes the cloud is
axisymmetric.  Metallicity has a negligible effect on the intrinsic
stellar luminosities, causing differences of only $\sim$0.03
magnitudes between 0.05Z$_{\odot}$ and 0.5Z$_{\odot}$, so we use a
0.1Z$_{\odot}$ model for our absolute magnitude calculations and
ignore any metallicity variations. Lastly, to account for the
uncertainty in the distance to the Smith Cloud, we recalculate the
allowed range of apparent OB star brightnesses in steps of 150 pc
($\sim 325/2$) from 12.4-1.3 to 12.4+1.3~kpc.  Over the full range of
possible distances, the minimum and maximum allowed apparent
magnitudes shift by $\sim$0.5~mags, but this shift does not have a
significant effect on the final results (see \S~\ref{sec:scob}).  The
accepted ranges of apparent magnitudes for OB stars in the Smith Cloud
are given in Table~\ref{table:magcuts}.  As with our
  color cuts, we allow a star to pass a brightness cut if its
  uncertainty potentially places it within the range of brightnesses
  occupied by OB stars.

 At this point, we can revisit the 45 per cent of NUV
  sources that are not detected by $WISE$.  Most of these objects have
  NUV magnitudes of 23--23.5, and to be undetected by $WISE$ they must
  have $W1 \ga 17$, the 5$\sigma$ sensitivity of the survey.  Our
  model apparent magnitudes show that objects this dim are
  inconsistent with OB stars at the distance of the Smith Cloud.
  Thus, we are at no risk of missing any candidates by ignoring these
  sources.  Additionally, our model magnitudes show that shallow
$GALEX$ AIS imaging is capable of detecting young OB stars at the
distance of the Smith Cloud.  The dimmest OB star we include has
m$_{\rm NUV}=18.2$, corresponding to $m_{\rm NUV}=20.2$ after adding
in the typical foreground extinction of $\sim$2 magnitudes.  Thus, the
dimmest OB star before extinction corrections is still brighter than
the AIS completeness limit of 20.5~magnitudes.

\begin{table}
\caption{Apparent Magnitude Limiting Values}
\label{table:magcuts}
\begin{tabular}{@{}lcc}
\hline
Band & Minimum & Maximum\\
\hline
$GALEX$ NUV & 8.1(7.8,8.3) & 18.2(17.9,18.4) \\
$WISE$ $W1$ & 10.7(10.5,11.0) & 16.8(16.5,17.0) \\
2MASS $J$ & 10.5(10.2,10.7) & 16.8(16.5,17.0) \\
\hline
\end{tabular}

\medskip
Apparent magnitudes assume a distance of 12.4~kpc and minimum/maximum
values reflect expected variation due to the intrinsic range of OB
star luminosities and the thickness of the Smith Cloud. Values in
parentheses give the expected range of magnitudes at the closest
(11.1~kpc) and furthest (13.7~kpc) possible Smith Cloud distances
based on the error bars from the existing distance estimate of
\citet{Lockman08}.  $GALEX$ magnitudes are given in the AB system,
while $WISE$ and 2MASS magnitudes are given in the Vega system.
\end{table}



\section{Results}
\label{sec:results}

\subsection{Number of Candidate OB Stars}
\label{sec:numob}

\begin{figure*}
\includegraphics[width=175mm]{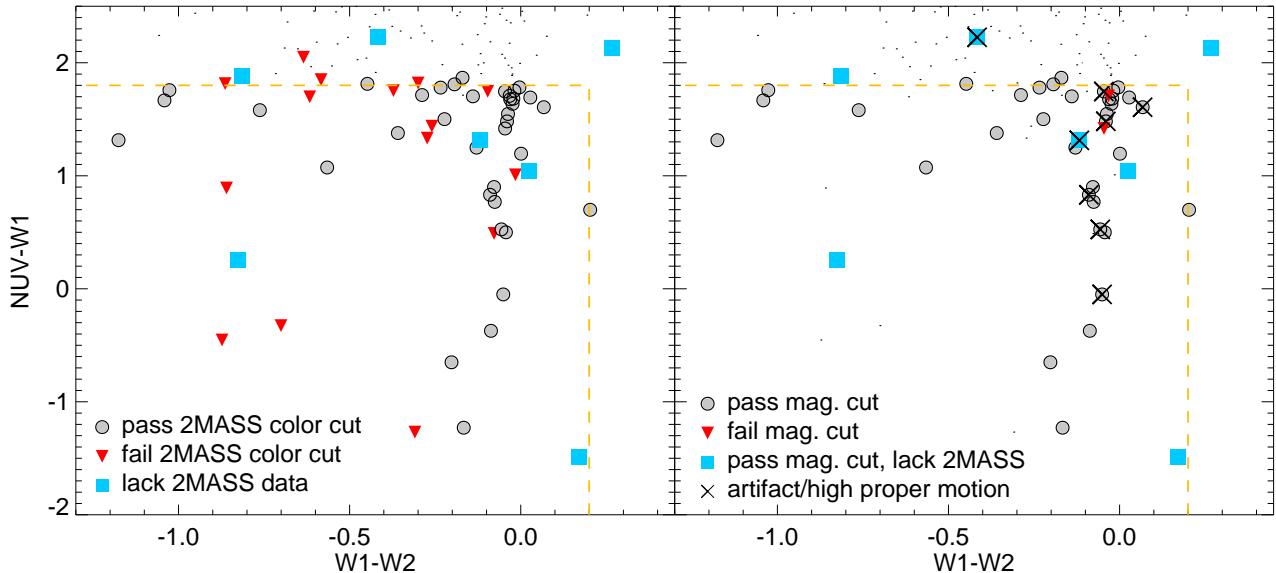}
\caption{Zoomed in version of Fig.~\ref{fig:synthetic} showing the
  results of the color and magnitude cuts. The region within the
  orange dashed line is identical to the hashed region in
  Fig.~\ref{fig:synthetic}. {\it (left)} Results of the color cuts
  described in \S~\ref{sec:colorcuts}.  All large points inside passed
  the NUV$-W1$ and $W1-W2$ color cuts.   Points lying outside the
  parameter space of OB stars can still pass the color cuts if their
  1$\sigma$ uncertainties potentially place them within the OB star region.
  Filled grey points represent stars that also passed the $J-H$ and
  $H-K$ cuts using 2MASS data.  Red triangles represent stars that
  failed the $J-H$ and $H-K$ cuts.  Blue filled squares represent
  stars that but lacked 2MASS data.  The grey and blue points are kept
  and passed through the magnitude cuts. {\it (right)} Results of the
  magnitude cuts described in \S~\ref{sec:magcuts}.  Stars that failed
  the previous color cuts are shown as small black dots.  Filled grey
  points represent stars that passed the NUV, $W1$, and $J$ band
  magnitude cuts assuming the Smith Cloud is at a distance of
  12.4~kpc.  Red triangles represent stars that failed the magnitude
  cuts in at least one band.  Blue points represent stars that passed
  the NUV and $W1$ magnitude cuts but lacked 2MASS data.  The X's
  denote stars that were cut due to having high proper motions or were
  potentially affected by $GALEX$ NUV image artifacts.  The 34 grey
  and blue points without X's make up our final sample of candidate OB
  stars shown in Fig.~\ref{fig:obcandidates}.}
\label{fig:colormagcut}
\end{figure*}

All sources in the Smith Cloud and the control fields are passed through
our calibrated color and apparent magnitude cuts to identify stars
that may plausibly be OB stars at the distance of the Smith Cloud.
Here we describe the results.

\subsubsection{Smith Cloud Field}
\label{sec:scob}

After applying $W1-W2$, NUV$-W1$, $J-H$, and $H-K$ color cuts, we are
left with 44 candidate OB stars in the Smith Cloud
field, seven of which lack 2MASS magnitudes
(Fig.~\ref{fig:colormagcut}).  Assuming the nominal distance of the
Smith Cloud, the apparent magnitude cuts reject two stars.  As
described in \S~\ref{sec:magcuts}, to see how the results depend on
the uncertainty in the distance to the cloud, we rerun the magnitude
cuts in steps from 11.1~kpc to 13.7~kpc.   Across the full
  range of distances, the total number of stars rejected by the
  magnitude cuts remains roughly constant at two stars, but up to 3--4
  stars are rejected close to the nearest and furthest allowed
  distances. Thus, after the magnitude cut, we are left with 40--42
  candidate OB stars, depending on the assumed distance to the Smith
  Cloud.

We then searched the SIMBAD stellar database to see if any of the
remaining stars had measured proper motions.  We used these proper
motions to calculate lower limits on their true space velocities
assuming the stars are at the distance of the Smith Cloud, and then
compared these velocities to the true motion of the Smith Cloud through
space, which is estimated to be $\sim$300~km~s$^{-1}$
\citep{Lockman08}.  Of the candidate stars, four have cataloged
proper motions, three of which yield velocities $\sim$1000~km~s$^{-1}$
or larger, which we judge to be unrealistically high and reject.  The
remaining star has a velocity $>$~70~km~s$^{-1}$.  We deem this to be
plausibly associated with the Smith Cloud and keep it in the list of
OB star candidates.

After accounting for measured proper motions, we examined all artifact
flags in the $GALEX$ catalog, and an additional  five
stars were rejected due to their fluxes being potentially corrupted by
NUV artifacts.  The combination of rejecting stars with high proper
motions and those potentially affected by NUV artifacts
serendipitously removes any dependence of our final result on the
distance to the cloud.  As described above, the magnitude cuts reject
slightly different numbers of stars depending on the assumed distance.
 Of the stars that were rejected at some distances but not
  others, all of them either had high proper motion or were affected
  by NUV artifacts, so ended up being rejected regardless of their
  apparent magnitudes.  In summary, we are left with a total of 34
  candidate OB stars found in the Smith Cloud field, 21 of which fall
  directly on the cloud.

\begin{figure}
\includegraphics[width=84mm]{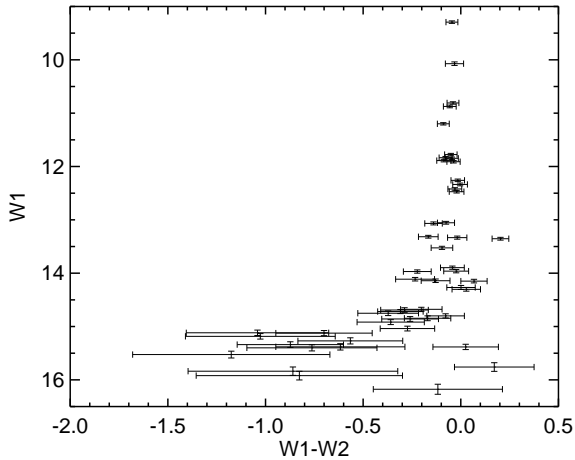}
\caption{ $W1$ vs. $W1-W2$ for all stars in the Smith Cloud field with
  NUV-$W1<1.8$.  At dimmer magnitudes, colors tend to be biased
  towards bluer values in conjunction with larger photometric
  uncertainties.}
\label{fig:w1w1w2}
\end{figure}

 Several of our final candidate OB stars lie at much bluer
  $W1-W2$ colors than expected for main sequence stars (which
  typically have $W1-W2\sim0$; see \S~\ref{sec:colorcuts}).
  Fig.~\ref{fig:w1w1w2} shows a $WISE$ color-magnitude diagram for all
  stars with NUV$-W1<1.8$.  Stars at dimmer magnitudes tend to be
  biased towards bluer colors.  Thus, we consider objects with such
  blue $W1-W2$ colors as valid OB star candidates, and attribute their
  unusual colors to large photometric uncertainties.

\begin{figure}
\includegraphics[width=84mm]{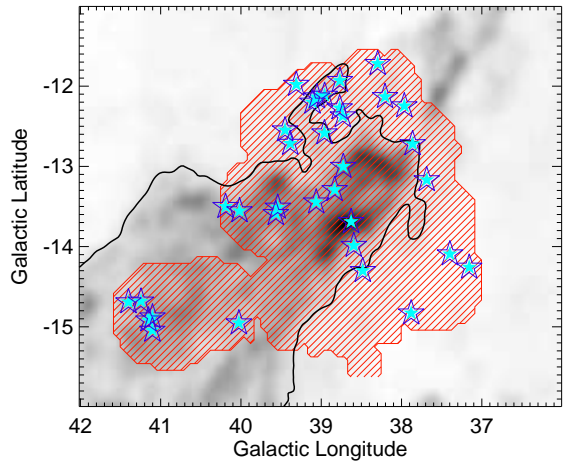}
\caption{The final 34 candidate OB stars (blue) overlaid on the Smith Cloud HI column density map.  The red filled region indicates the spatial extent of our $GALEX$ data, while the solid black line indicates the $5\times 10^{19}$~cm$^{-2}$ contour used to define the edge of the cloud.}
\label{fig:obcandidates}
\end{figure}

\subsubsection{Control Field}
\label{sec:controlob}

In the control field, the $GALEX$, $WISE$, and 2MASS color cuts leave 11 candidate stars, all of which have 2MASS magnitudes.
All of these stars pass the apparent magnitude cuts over the entire
range of possible distances to the Smith Cloud.  An additional two
stars are removed for having large proper motions, and
  three more stars are removed due to potential contamination from
NUV artifacts.  This selection leaves six candidate OB
stars in the control field.

\begin{table}
\caption{OB Star Candidates in Different Regions}
\label{table:stardens}
\begin{tabular}{@{}lccc}
\hline
Region & N & Area (kpc$^2$) & Density (kpc$^{-2}$)\\
\hline
On Cloud             & 21$\pm$4.6 & 0.28 & 75.0$\pm$16.4\\
Off Cloud            & 13$\pm$3.6 & 0.23 & 56.5$\pm$15.7\\
On+Off Cloud         & 35$\pm$5.9 & 0.51 & 68.6$\pm$11.5\\
Control              &  6$\pm$2.4 & 0.16 & 37.5$\pm$15.0\\
On Cloud $-$ Control & 11$\pm$6.2 & 0.28 & 39.3$\pm$22.1\\
Leo P                &  7$\pm$2.7 & 0.39 & 17.9$\pm$6.9 \\
\hline
\end{tabular}

\medskip
We adopt a distance of 12.4~kpc \citep{Lockman08} to calculate the physical area of all regions except Leo P, for which we adopt a distance of 1.75~Mpc \citep{Giovanelli13}.
\end{table}


\subsection{Likelihood of Association}
\label{sec:likelihood}

The numbers of stars and projected space densities in each region are
presented in Table~\ref{table:stardens}.  The uncertainties in
Table~\ref{table:stardens} are computed using Poisson statistics so that the
uncertainty of each measurement of $N$ stars is $\sqrt{N}$.  As
discussed before, there does not appear to be any additional
uncertainty introduced due to the uncertainty in the distance to the
Smith Cloud or its metallicity.

The 34 final candidate OB stars in the Smith Cloud field are overlaid
on the HI column density map in Fig.~\ref{fig:obcandidates} and the
surface densities of OB candidate stars in each region are compared in
Fig.~\ref{fig:stardens}.  Stars are preferentially found in the region
we have defined to be `on' the cloud, but the number of stars found
`on' versus `off' the cloud are the same within their uncertainties.
 Several stars are positioned very close to the cloud
  border, so that increasing the limiting column density by only
  0.1~dex would cause six more stars to fall in the `off' region,
  making the density of stars `off' the cloud larger than `on',
  although these numbers are still consistent within their
  uncertainties (lighter points in Fig.~\ref{fig:stardens}).

Instead of separating the Smith Cloud field into regions `on' and
`off' the cloud, we could justifiably consider the possibility that
all stars in this region are associated with the Smith Cloud.  This
assumption is reasonable considering stars can drift away from their
birth clouds over the course of their lifetimes.  Observations of
young clusters show typical radial velocity dispersions of
$\sim$20~km~s$^{-1}$ (see, e.g., the study of the Perseus OB2
association by \citealt{Steenbrugge03}).  Multiplying this by the
typical age of a B star ($\sim$100~Myr), drift distances of up to
2~kpc are possible over OB star lifetimes, equivalent to 9.4~degrees
at the distance of the Smith Cloud.  This angular radius encompasses
$>$~16 times the area of all our $GALEX$ coverage on the Smith Cloud.

Regardless of whether we assume all stars in our $GALEX$ fields are
associates with the cloud, or only those within the N$_{\rm HI} =
5\times 10^{19}$~cm$^{-2}$ contour, the spatial density of stars is
higher than the density of stars in the control field, but not at a
statistically significant level (Fig.~\ref{fig:stardens}).

\section{Discussion}
\label{sec:discussion}

\subsection{Comparison to Theoretical Expectations}
\label{sec:theory}

It is instructive to ask how many OB stars we might actually expect to
form in an object such as the Smith Cloud.  To address this question,
we measure the typical gas surface density in the cloud and use it to
infer the range of possible star formation rate surface densities.  In
regions within the outermost N$_{\rm HI}=5 \times 10^{19}$~cm$^{-2}$
contour and overlapping our search fields, the median column density
is $1.5\times 10^{20}$~cm$^{-2}$ or 1.2~M$_{\odot}$~pc$^{-2}$ (not
corrected for Helium).  \citet{Bigiel10b} examined the relationship
between star formation rate and HI surface density (also without
Helium corrections) on sub-kpc scales in the outer disks of dwarf
galaxies, and found that 75 per cent of regions at comparable gas
density form stars at rates between approximately
10$^{-(4.2-5.2)}$~M$_{\odot}$~yr$^{-1}$~kpc$^{-2}$.  We use this range
of star formation rates to calculate the number of OB stars at masses
greater than 2.15~M$_{\odot}$ (corresponding to stars $>10^4$~K
according to the same tabulations of stellar properties referenced in
\S~\ref{sec:magcuts}) expected under two scenarios: (1) one where the
Smith Cloud has been forming stars continuously for 10~Gyr, and (2)
another where the Smith Cloud has been forming stars only since its
first impact with the Milky Way 70~Myr ago
\citep{Lockman08,Nichols09}.  In both cases, we assume that the star
formation follows a Salpeter IMF, that the star formation rate and
average HI surface density have remained constant, and that stellar
ages can be approximated by
$\tau_{MS}=10^{10}(M/M_{\odot})^{-2.5}$~yr.

\begin{figure}
\includegraphics[width=84mm]{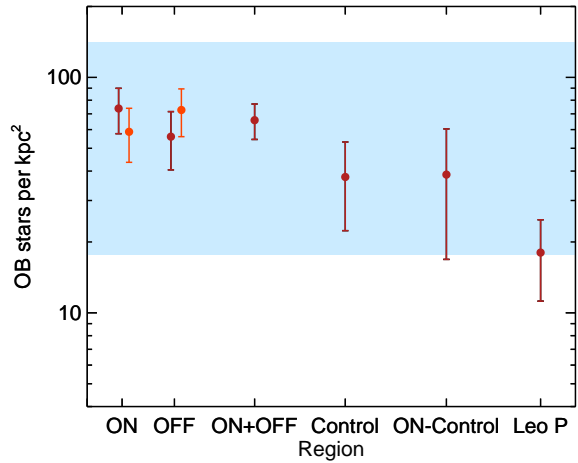}
\caption{The measured density of candidate OB stars in different
  regions of this study.  The blue band represents the range in
  surface density of OB stars given the range in possible star
  formation rates based on the typical gas density of the cloud, and
  assuming star formation began 70~Myr ago when the cloud first passed
  through the Galactic disk.  The lighter points immediately to the
  right of the `ON' and `OFF' points represent the values if the HI
  column density used to define the edge of the cloud is increased by
  0.1~dex.}
\label{fig:stardens}
\end{figure}

Under the 10~Gyr scenario, we expect between 45 and 390 OB stars on
the `ON' field of the Smith Cloud (or $\sim$160-1380~kpc$^{-2}$), far
more than we observe.  Under the 70~Myr scenario, we would expect to
find between 5 and 40 OB stars (or $\sim$18-140~kpc$^{-2}$) in this
same region, which is consistent with the observations.  The grey
region in Fig.~\ref{fig:stardens} shows the range of OB star densities
we would expect to measure in the 70~Myr scenario.  Unfortunately,
this range does not allow us to reject or confirm the Smith Cloud as a
star forming cloud because all the regions considered, including the
control field, are consistent with this star formation scenario.

For comparison, instead of calculating the expected number of OB stars
based on the star formation rate estimated using gas density, we can
reverse the calculation to estimate what star formation rate would
yield our observed number of stars, again assuming constant star
formation over either the last 10~Gyr or 70~Myr and that the star
formation follows a Salpeter IMF.  First, we assume all 21 candidate
OB stars on the cloud are indeed associated with the cloud.  To
recreate our observations under the 10~Gyr scenario would require a
star formation rate surface density of
 10$^{-5.5\pm0.1}$~M$_{\odot}$~yr$^{-1}$~kpc$^{-2}$.  Under the 70~Myr
scenario, a star formation rate surface density of
 10$^{-4.5\pm0.1}$~M$_{\odot}$~yr$^{-1}$~kpc$^{-2}$ would create the
observed number of stars.  Alternatively, if we proceed with the
assumption that the OB star density measured in the control field is a
good measure of the background from the Milky Way, then 10 out of the
17 observed stars are actually associated with the Smith Cloud.  This
implies star formation rate surface densities of
 $10^{-5.8\pm0.3}$~yr$^{-1}$~kpc$^{-2}$ and
$10^{-4.8\pm0.3}$~yr$^{-1}$~kpc$^{-2}$ for the 10~Gyr and 70~Myr
scenarios respectively.  Whether we subtract a background OB star
density or not, the 70~Myr scenario is the only one consistent with
the expected range of star formation rates based on gas surface
density from \citet{Bigiel10b}, and is likely a better description of
the cloud's star formation history.

\subsection{Comparison with Leo P}
\label{sec:leop}

The newly discovered galaxy Leo P was originally identified as a UCHVC
within the ALFALFA survey \citep{Giovanelli10}.  Follow-up
observations by \citet{McQuinn13} and \cite{Rhode13} revealed the
clear presence of a stellar population and ongoing star formation,
making it the best existing example of an `HVC turned galaxy'.
Since we are trying to determine whether the Smith Cloud is a similar
object, it is instructive to compare its properties with Leo P.

Despite its original designation as a UCHVC, Leo P is not dramatically
smaller than the Smith Cloud.  HI observations with the JVLA and
Arecibo reveal that it is approximately 1/3 the size of the Smith
Cloud and has almost an equivalent HI mass \citep{Giovanelli13}.
However, given the Smith Cloud's recent history of interaction with
the Milky Way, its original HI mass may have been much larger
($\sim$10~times greater according to \citealt{Nichols09}), and it has
an ionized gas component equal to or greater than its neutral
component \citep{Hill09}.  The ionized gas fraction of Leo~P is
unconstrained.  The peak HI column densities of the Smith Cloud and
Leo P are very comparable.   At an equivalent spatial
  resolution of $\sim$30~pc, both show comparable peak column
  densities around $\sim5\times10^{20}$~cm$^{-2}$
  \citep{Lockman08,Bernstein-Cooper14}.  Leo P has a gas phase
metallicity of $\sim$3\%~Z$_{\odot}$ \citep{Skillman13}, consistent
with (or possibly even lower than) that of the Smith Cloud.

\begin{figure}
\includegraphics[width=84mm]{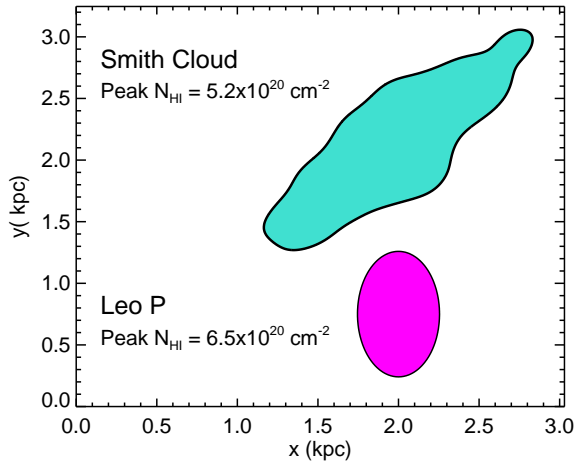}
\caption{Leo P's size relative to that of the Smith Cloud if both were
  placed at the same distance.  Outlines represent the
  5$\times10^{19}$~cm$^{-2}$ contours.  The Smith Cloud outline is
  based on an HI map convolved to 200 pc resolution.  The Leo~P shape is
  an approximation based on the published radius and axis ratio from
  \citet{Giovanelli13}.  Peak column densities come from \citet{Lockman08} and \citet{Bernstein-Cooper14}.} 
\label{fig:sc_leop}
\end{figure}

Deep $V$ and $I$ band imaging of Leo~P was presented in
\citet{McQuinn13}.  We use the $V-I$ color-temperature relation of
\citet{Worthey11} to isolate OB stars with surface temperatures
$>10^4$~K, or $V-I < 0.036$.  Using this color cut, we find a total of
7 OB stars in Leo P.  Like we did for the Smith Cloud, if we use the approximate area within the 5$\times 10^{19}$~cm$^{-2}$ contour (which contains all the identified OB stars,
although the photometric data extend to larger radius), we obtain an OB
star density less than that found in the direction of the Smith Cloud
(see Fig.~\ref{fig:stardens}, Table~\ref{table:stardens}).
Therefore, if Leo P were placed at the position of the Smith Cloud,
its OB star population would not be detected to a statistically
significant level by our analysis.  This result implies that star
formation within the Smith Cloud may be present but simply undetected
due to Milky Way contamination.

\section{Conclusions}
\label{sec:conclusion}

Using a combination of 21cm, NIR, and UV data, we conducted a search
for young stars potentially associated with the Smith Cloud, an HVC
currently on a collision course with the Milky Way.  Using a
combination of color and apparent magnitude cuts, we identified 34
candidate OB stars out of $\sim$40,000 sources in a field overlapping
the Smith cloud, with 21 of these stars falling directly over the
cloud, although this number is sensitive to the specific
definition of the cloud edge.  Regardless, the number density of
candidate stars in the Smith Cloud field is slightly higher than the
density of a control region, but not at a statistically significant
level.

We have compared the number of candidate OB stars to theoretical
expectations based on the gas density of the cloud.  We find that the
number of candidate stars is consistent with a scenario where constant
star formation began 70~Myr ago when the cloud first impacted the
Galactic disk, but is not consistent with a scenario where the cloud has
been constantly forming stars over the last several Gyr.  We estimate
the star formation rate surface density of the cloud to be
$10^{-4.8\pm0.3}$~M$_{\odot}$~yr$^{-1}$~kpc$^{-2}$, again assuming
constant star formation triggered $\sim$70~Myr ago.

 Lastly, we have compared the properties of the Smith
  Cloud with those of the recently discovered star forming galaxy that
  was originally classified as a UCHVC, Leo~P.  We find that these two
  objects have comparable HI properties (mass, size, peak column
  density) and gas-phase metallicities.  Leo~P, although hosting a
  stellar population and ongoing star formation, would not be
  detectable via our method if it were placed at the position of the
  Smith Cloud.  This means that if the Smith Cloud is forming stars,
  it may be doing so at a low rate comparable to dwarf galaxies like
  Leo~P and yet be indiscernible from the Milky Way foreground using
  our photometric techniques alone.

On the other hand, if the Smith Cloud is not forming stars, this
raises the question of how it is different from Leo P.  Both objects
show comparable maximum column densities, but Leo P's column density
may translate into a higher {\it volume} density since the gas is
confined to a disk, whereas the Smith Cloud has an unknown thickness
along the line of sight.  Any lack of star formation in the Smith
Cloud would be unlikely to be due to metallicity considering that
Leo~P's metallicity is comparable to the Smith Cloud's.  A final major
consideration is the difference in dynamical properties of these
systems. The Smith Cloud is undergoing a major interaction with the
Milky Way.  While the tidal forces or cloud collisions involved in
this interaction could promote star formation (initially, we thought
this interaction might raise the likelihood of seeing star formation),
the cloud's disruption may impart enough additional turbulence to help
support against collapse, all while gas is continuously being removed
via stripping by the corona \citep{Heitsch09}.  Meanwhile, Leo~P is
relatively isolated with no known object within 0.5 Mpc
\citep{McQuinn13}.  However, any effect attributed to these
differences is at this point speculative, as our data are consistent
with the possibility that the Smith Cloud does host star formation.

Our results leave open the possibility that the Smith Cloud is a
galaxy yet waiting to be discovered.  Unfortunately, its position with
respect to the Milky Way and its vast extent make studying it
challenging.  Spectroscopy of the stars at the position of the Smith
Cloud, to identify distinct kinematics or metallicities, is a
potentially promising path to determine whether it hosts its own
stellar population.

\section*{Acknowledgments}

We thank the referee for their helpful comments that improved this
paper.  We also thank Jay Lockman for sharing his HI data cube for the
Smith Cloud and instructing us on its use, as well as for his helpful
comments for this study.  We also thank Fabian Heitsch for his
valuable input on the $GALEX$ proposal from which we obtained our data
and the overall study.  We also thank Alex Hill and Mary Putman for
useful discussions and suggestions.  This program was funded by the
$GALEX$ Guest Investigator program under NASA grant NNX10AR88G.  This
publication makes use of data products from the Wide-field Infrared
Survey Explorer, which is a joint project of the University of
California, Los Angeles, and the Jet Propulsion Laboratory/California
Institute of Technology, funded by the National Aeronautics and Space
Administration.  This publication makes use of data products from the
Two Micron All Sky Survey, which is a joint project of the University
of Massachusetts and the Infrared Processing and Analysis
Center/California Institute of Technology, funded by the National
Aeronautics and Space Administration and the National Science
Foundation.  This research has made use of the SIMBAD database,
operated at CDS, Strasbourg, France.

\bibliographystyle{mn2e}
\bibliography{mybib}

\begin{thebibliography}{57}
\expandafter\ifx\csname natexlab\endcsname\relax\def\natexlab#1{#1}\fi

\bibitem[{{Adams} {et~al}\mbox{.}(2013){Adams}, {Giovanelli}, \&
  {Haynes}}]{Adams13}
{Adams} E.~A.~K., {Giovanelli} R., {Haynes} M.~P., 2013, \apj, 768, 77

\bibitem[{{Bernstein-Cooper} {et~al}\mbox{.}(2014){Bernstein-Cooper}, {Cannon},
  {Elson}, {Warren}, {Chengular}, {Skillman}, {Adams}, {Bolatto}, {Giovanelli},
  {Haynes}, {McQuinn}, {Pardy}, {Rhode}, \& {Salzer}}]{Bernstein-Cooper14}
{Bernstein-Cooper} E.~Z. {et~al.}, 2014, \aj, 148, 35

\bibitem[{{Bigiel} {et~al}\mbox{.}(2010){Bigiel}, {Leroy}, {Walter}, {Blitz},
  {Brinks}, {de Blok}, \& {Madore}}]{Bigiel10b}
{Bigiel} F., {Leroy} A., {Walter} F., {Blitz} L., {Brinks} E., {de Blok}
  W.~J.~G., {Madore} B., 2010, \aj, 140, 1194

\bibitem[{{Bigiel} {et~al}\mbox{.}(2008){Bigiel}, {Leroy}, {Walter}, {Brinks},
  {de Blok}, {Madore}, \& {Thornley}}]{Bigiel08}
{Bigiel} F., {Leroy} A., {Walter} F., {Brinks} E., {de Blok} W.~J.~G., {Madore}
  B., {Thornley} M.~D., 2008, \aj, 136, 2846

\bibitem[{{Bland-Hawthorn} {et~al}\mbox{.}(1998){Bland-Hawthorn}, {Veilleux},
  {Cecil}, {Putman}, {Gibson}, \& {Maloney}}]{BlandHawthorn98}
{Bland-Hawthorn} J., {Veilleux} S., {Cecil} G.~N., {Putman} M.~E., {Gibson}
  B.~K., {Maloney} P.~R., 1998, \mnras, 299, 611

\bibitem[{{Blitz} {et~al}\mbox{.}(1999){Blitz}, {Spergel}, {Teuben},
  {Hartmann}, \& {Burton}}]{Blitz99}
{Blitz} L., {Spergel} D.~N., {Teuben} P.~J., {Hartmann} D., {Burton} W.~B.,
  1999, \apj, 514, 818

\bibitem[{{Braun} \& {Burton}(1999)}]{Braun99}
{Braun} R., {Burton} W.~B., 1999, \aap, 341, 437

\bibitem[{{Casetti-Dinescu} {et~al}\mbox{.}(2014){Casetti-Dinescu}, {Moni
  Bidin}, {Girard}, {M{\'e}ndez}, {Vieira}, {Korchagin}, \& {van
  Altena}}]{Casetti14}
{Casetti-Dinescu} D.~I., {Moni Bidin} C., {Girard} T.~M., {M{\'e}ndez} R.~A.,
  {Vieira} K., {Korchagin} V.~I., {van Altena} W.~F., 2014, \apjl, 784, L37

\bibitem[{{Christodoulou} {et~al}\mbox{.}(1997){Christodoulou}, {Tohline}, \&
  {Keenan}}]{Christodoulou97}
{Christodoulou} D.~M., {Tohline} J.~E., {Keenan} F.~P., 1997, \apj, 486, 810

\bibitem[{{de Blok} \& {Walter}(2006)}]{deBlok06}
{de Blok} W.~J.~G., {Walter} F., 2006, \aj, 131, 363

\bibitem[{{Giovanelli} {et~al}\mbox{.}(2013){Giovanelli}, {Haynes}, {Adams},
  {Cannon}, {Rhode}, {Salzer}, {Skillman}, {Bernstein-Cooper}, \&
  {McQuinn}}]{Giovanelli13}
{Giovanelli} R. {et~al.}, 2013, \aj, 146, 15

\bibitem[{{Giovanelli} {et~al}\mbox{.}(2010){Giovanelli}, {Haynes}, {Kent}, \&
  {Adams}}]{Giovanelli10}
{Giovanelli} R., {Haynes} M.~P., {Kent} B.~R., {Adams} E.~A.~K., 2010, \apjl,
  708, L22

\bibitem[{{Gould} \& {Salpeter}(1963)}]{Gould63}
{Gould} R.~J., {Salpeter} E.~E., 1963, \apj, 138, 393

\bibitem[{{Habets} \& {Heintze}(1981)}]{Habets81}
{Habets} G.~M.~H.~J., {Heintze} J.~R.~W., 1981, \aaps, 46, 193

\bibitem[{{Heitsch} \& {Putman}(2009)}]{Heitsch09}
{Heitsch} F., {Putman} M.~E., 2009, \apj, 698, 1485

\bibitem[{{Hernandez} {et~al}\mbox{.}(2013){Hernandez}, {Wakker}, {Benjamin},
  {French}, {Kerp}, {Lockman}, {O'Toole}, \& {Winkel}}]{Hernandez13}
{Hernandez} A.~K., {Wakker} B.~P., {Benjamin} R.~A., {French} D., {Kerp} J.,
  {Lockman} F.~J., {O'Toole} S., {Winkel} B., 2013, \apj, 777, 19

\bibitem[{{Hill} {et~al}\mbox{.}(2009){Hill}, {Haffner}, \&
  {Reynolds}}]{Hill09}
{Hill} A.~S., {Haffner} L.~M., {Reynolds} R.~J., 2009, \apj, 703, 1832

\bibitem[{{Hill} {et~al}\mbox{.}(2013){Hill}, {Mao}, {Benjamin}, {Lockman}, \&
  {McClure-Griffiths}}]{Hill13}
{Hill} A.~S., {Mao} S.~A., {Benjamin} R.~A., {Lockman} F.~J.,
  {McClure-Griffiths} N.~M., 2013, \apj, 777, 55

\bibitem[{{Hopp} {et~al}\mbox{.}(2003){Hopp}, {Schulte-Ladbeck}, \&
  {Kerp}}]{Hopp03}
{Hopp} U., {Schulte-Ladbeck} R.~E., {Kerp} J., 2003, \mnras, 339, 33

\bibitem[{{Hopp} {et~al}\mbox{.}(2007){Hopp}, {Schulte-Ladbeck}, \&
  {Kerp}}]{Hopp07}
{Hopp} U., {Schulte-Ladbeck} R.~E., {Kerp} J., 2007, \mnras, 374, 1164

\bibitem[{{Ivezic} \& {Christodoulou}(1997)}]{Ivezic97}
{Ivezic} Z., {Christodoulou} D.~M., 1997, \apj, 486, 818

\bibitem[{{Joung} {et~al}\mbox{.}(2012){Joung}, {Bryan}, \& {Putman}}]{Joung12}
{Joung} M.~R., {Bryan} G.~L., {Putman} M.~E., 2012, \apj, 745, 148

\bibitem[{{Klypin} {et~al}\mbox{.}(1999){Klypin}, {Kravtsov}, {Valenzuela}, \&
  {Prada}}]{Klypin99}
{Klypin} A., {Kravtsov} A.~V., {Valenzuela} O., {Prada} F., 1999, \apj, 522, 82

\bibitem[{{Lejeune} {et~al}\mbox{.}(1998){Lejeune}, {Cuisinier}, \&
  {Buser}}]{Lejeune98}
{Lejeune} T., {Cuisinier} F., {Buser} R., 1998, VizieR Online Data Catalog,
  413, 65

\bibitem[{{Lockman} {et~al}\mbox{.}(2008){Lockman}, {Benjamin}, {Heroux}, \&
  {Langston}}]{Lockman08}
{Lockman} F.~J., {Benjamin} R.~A., {Heroux} A.~J., {Langston} G.~I., 2008,
  \apjl, 679, L21

\bibitem[{{Mac Low} \& {Glover}(2012)}]{MacLow12}
{Mac Low} M.-M., {Glover} S.~C.~O., 2012, \apj, 746, 135

\bibitem[{{Mace} {et~al}\mbox{.}(2013){Mace}, {Kirkpatrick}, {Cushing},
  {Gelino}, {Griffith}, {Skrutskie}, {Marsh}, {Wright}, {Eisenhardt}, {McLean},
  {Thompson}, {Mix}, {Bailey}, {Beichman}, {Bloom}, {Burgasser}, {Fortney},
  {Hinz}, {Knox}, {Lowrance}, {Marley}, {Morley}, {Rodigas}, {Saumon},
  {Sheppard}, \& {Stock}}]{Mace13}
{Mace} G.~N. {et~al.}, 2013, \apjs, 205, 6

\bibitem[{{Massey} {et~al}\mbox{.}(2004){Massey}, {Bresolin}, {Kudritzki},
  {Puls}, \& {Pauldrach}}]{Massey04}
{Massey} P., {Bresolin} F., {Kudritzki} R.~P., {Puls} J., {Pauldrach} A.~W.~A.,
  2004, \apj, 608, 1001

\bibitem[{{Massey} {et~al}\mbox{.}(2005){Massey}, {Puls}, {Pauldrach},
  {Bresolin}, {Kudritzki}, \& {Simon}}]{Massey05}
{Massey} P., {Puls} J., {Pauldrach} A.~W.~A., {Bresolin} F., {Kudritzki} R.~P.,
  {Simon} T., 2005, \apj, 627, 477

\bibitem[{{Massey} {et~al}\mbox{.}(2009){Massey}, {Zangari}, {Morrell}, {Puls},
  {DeGioia-Eastwood}, {Bresolin}, \& {Kudritzki}}]{Massey09}
{Massey} P., {Zangari} A.~M., {Morrell} N.~I., {Puls} J., {DeGioia-Eastwood}
  K., {Bresolin} F., {Kudritzki} R.-P., 2009, \apj, 692, 618

\bibitem[{{McQuinn} {et~al}\mbox{.}(2013){McQuinn}, {Skillman}, {Berg},
  {Cannon}, {Salzer}, {Adams}, {Dolphin}, {Giovanelli}, {Haynes}, \&
  {Rhode}}]{McQuinn13}
{McQuinn} K.~B.~W. {et~al.}, 2013, \aj, 146, 145

\bibitem[{{Morrissey} {et~al}\mbox{.}(2007){Morrissey}, {Conrow}, {Barlow},
  {Small}, {Seibert}, {Wyder}, {Budav{\'a}ri}, {Arnouts}, {Friedman},
  {Forster}, {Martin}, {Neff}, {Schiminovich}, {Bianchi}, {Donas}, {Heckman},
  {Lee}, {Madore}, {Milliard}, {Rich}, {Szalay}, {Welsh}, \&
  {Yi}}]{Morrissey07}
{Morrissey} P. {et~al.}, 2007, \apjs, 173, 682

\bibitem[{{Nichols} \& {Bland-Hawthorn}(2009)}]{Nichols09}
{Nichols} M., {Bland-Hawthorn} J., 2009, \apj, 707, 1642

\bibitem[{{Nichols} {et~al}\mbox{.}(2014){Nichols}, {Mirabal}, {Agertz},
  {Lockman}, \& {Bland-Hawthorn}}]{Nichols14}
{Nichols} M., {Mirabal} N., {Agertz} O., {Lockman} F.~J., {Bland-Hawthorn} J.,
  2014, ArXiv e-prints

\bibitem[{{O'Donnell}(1994)}]{Odonnell94}
{O'Donnell} J.~E., 1994, \apj, 422, 158

\bibitem[{{Putman} {et~al}\mbox{.}(2003){Putman}, {Bland-Hawthorn}, {Veilleux},
  {Gibson}, {Freeman}, \& {Maloney}}]{Putman03}
{Putman} M.~E., {Bland-Hawthorn} J., {Veilleux} S., {Gibson} B.~K., {Freeman}
  K.~C., {Maloney} P.~R., 2003, \apj, 597, 948

\bibitem[{{Putman} {et~al}\mbox{.}(2012){Putman}, {Peek}, \&
  {Joung}}]{Putman12}
{Putman} M.~E., {Peek} J.~E.~G., {Joung} M.~R., 2012, \araa, 50, 491

\bibitem[{{Repolust} {et~al}\mbox{.}(2004){Repolust}, {Puls}, \&
  {Herrero}}]{Repolust04}
{Repolust} T., {Puls} J., {Herrero} A., 2004, \aap, 415, 349

\bibitem[{{Rhode} {et~al}\mbox{.}(2013){Rhode}, {Salzer}, {Haurberg}, {Van
  Sistine}, {Young}, {Haynes}, {Giovanelli}, {Cannon}, {Skillman}, {McQuinn},
  \& {Adams}}]{Rhode13}
{Rhode} K.~L. {et~al.}, 2013, \aj, 145, 149

\bibitem[{{Richter} {et~al}\mbox{.}(2001){Richter}, {Sembach}, {Wakker}, \&
  {Savage}}]{Richter01}
{Richter} P., {Sembach} K.~R., {Wakker} B.~P., {Savage} B.~D., 2001, \apjl,
  562, L181

\bibitem[{{Schlegel} {et~al}\mbox{.}(1998){Schlegel}, {Finkbeiner}, \&
  {Davis}}]{Schlegel98}
{Schlegel} D.~J., {Finkbeiner} D.~P., {Davis} M., 1998, \apj, 500, 525

\bibitem[{{Sembach} {et~al}\mbox{.}(2001){Sembach}, {Howk}, {Savage}, \&
  {Shull}}]{Sembach01}
{Sembach} K.~R., {Howk} J.~C., {Savage} B.~D., {Shull} J.~M., 2001, \aj, 121,
  992

\bibitem[{{Siegel} {et~al}\mbox{.}(2005){Siegel}, {Majewski}, {Gallart},
  {Sohn}, {Kunkel}, \& {Braun}}]{Siegel05}
{Siegel} M.~H., {Majewski} S.~R., {Gallart} C., {Sohn} S.~T., {Kunkel} W.~E.,
  {Braun} R., 2005, \apj, 623, 181

\bibitem[{{Simon} \& {Blitz}(2002)}]{Simon02}
{Simon} J.~D., {Blitz} L., 2002, \apj, 574, 726

\bibitem[{{Skillman} {et~al}\mbox{.}(2013){Skillman}, {Salzer}, {Berg},
  {Pogge}, {Haurberg}, {Cannon}, {Aver}, {Olive}, {Giovanelli}, {Haynes},
  {Adams}, {McQuinn}, \& {Rhode}}]{Skillman13}
{Skillman} E.~D. {et~al.}, 2013, \aj, 146, 3

\bibitem[{{Smith}(1963)}]{Smith63}
{Smith} G.~P., 1963, \bain, 17, 203

\bibitem[{{Steenbrugge} {et~al}\mbox{.}(2003){Steenbrugge}, {de Bruijne},
  {Hoogerwerf}, \& {de Zeeuw}}]{Steenbrugge03}
{Steenbrugge} K.~C., {de Bruijne} J.~H.~J., {Hoogerwerf} R., {de Zeeuw} P.~T.,
  2003, \aap, 402, 587

\bibitem[{{Sternberg} {et~al}\mbox{.}(2002){Sternberg}, {McKee}, \&
  {Wolfire}}]{Sternberg02}
{Sternberg} A., {McKee} C.~F., {Wolfire} M.~G., 2002, \apjs, 143, 419

\bibitem[{{Strai{\v z}ys} \& {Lazauskait{\.e}}(2009)}]{Straizys09}
{Strai{\v z}ys} V., {Lazauskait{\.e}} R., 2009, Baltic Astronomy, 18, 19

\bibitem[{{Thilker} {et~al}\mbox{.}(2007){Thilker}, {Bianchi}, {Meurer}, {Gil
  de Paz}, {Boissier}, {Madore}, {Boselli}, {Ferguson}, {Mu{\~n}oz-Mateos},
  {Madsen}, {Hameed}, {Overzier}, {Forster}, {Friedman}, {Martin}, {Morrissey},
  {Neff}, {Schiminovich}, {Seibert}, {Small}, {Wyder}, {Donas}, {Heckman},
  {Lee}, {Milliard}, {Rich}, {Szalay}, {Welsh}, \& {Yi}}]{Thilker07}
{Thilker} D.~A. {et~al.}, 2007, \apjs, 173, 538

\bibitem[{{Wakker} \& {van Woerden}(1997)}]{Wakker97}
{Wakker} B.~P., {van Woerden} H., 1997, \araa, 35, 217

\bibitem[{{Wakker} {et~al}\mbox{.}(2008){Wakker}, {York}, {Wilhelm},
  {Barentine}, {Richter}, {Beers}, {Ivezi{\'c}}, \& {Howk}}]{Wakker08}
{Wakker} B.~P., {York} D.~G., {Wilhelm} R., {Barentine} J.~C., {Richter} P.,
  {Beers} T.~C., {Ivezi{\'c}} {\v Z}., {Howk} J.~C., 2008, \apj, 672, 298

\bibitem[{{Willman} {et~al}\mbox{.}(2002){Willman}, {Dalcanton}, {Ivezi{\'c}},
  {Schneider}, \& {York}}]{Willman02}
{Willman} B., {Dalcanton} J., {Ivezi{\'c}} {\v Z}., {Schneider} D.~P., {York}
  D.~G., 2002, \aj, 124, 2600

\bibitem[{{Worthey} \& {Lee}(2011)}]{Worthey11}
{Worthey} G., {Lee} H.-c., 2011, \apjs, 193, 1

\bibitem[{{Wright} {et~al}\mbox{.}(2010){Wright}, {Eisenhardt}, {Mainzer},
  {Ressler}, {Cutri}, {Jarrett}, {Kirkpatrick}, {Padgett}, {McMillan},
  {Skrutskie}, {Stanford}, {Cohen}, {Walker}, {Mather}, {Leisawitz}, {Gautier},
  {McLean}, {Benford}, {Lonsdale}, {Blain}, {Mendez}, {Irace}, {Duval}, {Liu},
  {Royer}, {Heinrichsen}, {Howard}, {Shannon}, {Kendall}, {Walsh}, {Larsen},
  {Cardon}, {Schick}, {Schwalm}, {Abid}, {Fabinsky}, {Naes}, \&
  {Tsai}}]{Wright10}
{Wright} E.~L. {et~al.}, 2010, \aj, 140, 1868

\bibitem[{{Wu} {et~al}\mbox{.}(2012){Wu}, {Hao}, {Jia}, {Zhang}, \&
  {Peng}}]{Wu12}
{Wu} X.-B., {Hao} G., {Jia} Z., {Zhang} Y., {Peng} N., 2012, \aj, 144, 49

\bibitem[{{Yan} {et~al}\mbox{.}(2013){Yan}, {Donoso}, {Tsai}, {Stern}, {Assef},
  {Eisenhardt}, {Blain}, {Cutri}, {Jarrett}, {Stanford}, {Wright}, {Bridge}, \&
  {Riechers}}]{Yan13}
{Yan} L. {et~al.}, 2013, \aj, 145, 55

\end{thebibliography}

\label{lastpage}

\end{document}